# Delineating complex ferroelectric domain structures via second harmonic generation spectral imaging

Wei Li[a], Yunpeng Ma[a], Tianyi Feng[a], Sergei V. Kalinin[b], Jing-Feng Li[a,**], Qian Li[a,*]

[a] *State Key Laboratory of New Ceramics and Fine Processing, School of Materials Science and Engineering, Tsinghua University, Beijing, 100084, China*
[b] *Department of Materials Science and Engineering, University of Tennessee Knoxville, Knoxville, TN 37996*

[*] Corresponding author.
[**] Corresponding author.
Email addresses: qianli_mse@tsinghua.edu.cn (Q. Li), jingfeng@mail.tsinghua.edu.cn (J.-F. Li).

**Abstract**

Understanding the mechanisms and spatial correlations of crystallographic symmetry breaking in ferroelectric materials is essential to tuning their functional properties. While optical second harmonic generation (SHG) has long been utilized in ferroelectric studies, its capability for probing complex polar materials has yet to be fully realized. Here, we develop a SHG spectral imaging method implemented on a home-designed laser-scanning SHG microscope, and demonstrate its application for a model system of $(K,Na)NbO_3$ single crystals. Supervised model fitting analysis produces comprehensive information about the polarization vector orientations and relative fractions of constituent domain variants as well as their thermal evolution across the polymorphic phase transitions. We observe an unexpected persistence of the orthorhombic phase at low temperatures, pointing to the phase competitions. Besides, we show that unsupervised matrix decomposition analysis can quickly and faithfully reveal domain configurations without a priori knowledge about specific material systems. The SHG spectral imaging method can be readily extended to other ferroelectric materials with potentials to be further enhanced.

**Keywords**

## 1. Introduction

Ferroelectric materials boast a century's history of research and development, enabling a wide range of industrial device applications, such as precision actuators, dielectric capacitors, nonvolatile memories and microwave filters [1–4]. Recent years have witnessed an exciting new tide of discoveries as exemplified by topological polar nanostructures, two-dimensional and fluorite-structure ferroelectric materials, which both expand our horizons of polar phenomena in materials and foster next-generation electronic technologies [5–7]. The study of ferroelectrics hinges on various characterization tools, among which optical second harmonic generation (SHG) has long played a significant role [8–11]. SHG is a two-photon, frequency-doubling nonlinear process that only occurs in materials without centrosymmetry, as described by **Eqn. 1**:

$$I_{2\omega} \propto P_{2\omega}^2 = \left( \sum d_{ij} E_i E_j \right)^2 \tag{1}$$

where $I_{2\omega}$ and $P_{2\omega}$ are the intensity and polarization strength, respectively, of second harmonic ($2\omega$) light, $E_j$ the electric field strength of fundamental ($1\omega$) light, and $d_{ij}$ nonlinear coefficients (third-rank tensor). For ferroelectrics, the crystallographic symmetry-breaking inevitably leads to the occurrence of SHG. The latter in turn engenders a method to determine the structural characteristics of materials [12–14]. SHG can be highly sensitive to local polar distortions by virtue of high-power ultrafast laser excitation, and is essentially independent with charge carrier transport processes (which can plague low-frequency electrical measurements as for those leaky materials) [15,16]. These advantages of SHG, supplemented with convenience of non-contact and non-invasiveness, have established itself a decisive tool for the identification of new polar materials in both bulk and thin-film forms [17–19].

The prevailing SHG measurement is tracking the response signals while varying the relative polarization directions of the $1\omega$ excitation light and $2\omega$ light analyzer (that is, polarimetry), thereby to determine the point group symmetries of the materials. SHG polarimetry is particularly useful when coupled with in-situ temperature, electric field and other stimuli, revealing variations associated with the phase transition behaviors [20–23]. For the purpose of spatially-resolved measurements, SHG microscopy has also been available as operated in either a scanning mode or a full-field mode [24]. In the scanning mode, the laser beam is tightly focused (often to a sub-micron, diffraction limit size) and raster-scanned onto the sample to form images point-by-point; the full-field mode forms images through an objective lens over a large simultaneously illuminated region and thus is more stringent on sample responsivity (since SHG intensity scales with the square of excitation light power density). SHG microscopy has been applied to image mesoscopic heterogeneities in ferroelectric materials, for example, domain structures [25,26], non-Ising domain walls [27,28] and chiral textures [29].

Despite all the success achieved thus far, the amount of information on materials functionality from SHG has so far been limited. SHG polarimetry and microscopy provide spectral and imaging information, respectively; however, their combination is rarely implemented so far. More specifically, acquiring a set of SHG spectra in a spatially-resolved manner can obtain abundant information devoid of measurement biases and provide a basis for comprehensive analyses of local polar symmetries and other types of dependencies. This may be especially significant for probing complex polar materials where multiple phases coexist along with correlated local distortions.

In this study, we have developed a SHG spectral imaging method based on efficient acquisition of polarization angle-dependent image datasets as facilitated by a home-designed laser-scanning SHG microscope. We introduce both supervised model fitting and unsupervised big-data approaches to analyze the spectral imaging datasets. We apply this method to a model system of $K_{0.5}Na_{0.5}NbO_3$ (KNN) single crystals, one of the most promising lead-free piezoelectric materials [30,31]. The structural characteristics of KNN features a complex structural transition sequence from rhombohedral (*R* phase, space group: *R3c*) to orthorhombic (*O* phase, *Amm2*) and then from tetragonal (*T* phase, *P4mm*) to cubic phase (*Pm-3m*) on heating [1]. The stabilities of these polymorphic phases can be tuned by chemical doping to enhance the piezoelectric properties, whereas the associated multi-phase domain structures often pose challenges for characterization at quantitative levels [32-35]. Our results clearly delineate the spatial configurations and temperature evolution behaviors of domains in KNN with renewed insights, thus demonstrating the applicability of SHG spectral imaging methods.

## 2. Methods

### 2.1. Sample preparation

The KNN crystal sample was prepared by top-seeded solution growth (TSSG) method. Powders of $K_2CO_3$, $Na_2CO_3$, and $Nb_2O_5$ were weighed to obtain a composition of $K_{1-x}Na_xNbO_3$ and were melted in a platinum crucible with excess $K_2CO_3$ and $Na_2CO_3$ as self-flux. The velocities of rotation and pulling of the crystals during crystal growth were ~6 rpm and ~0.5 mm per hour, respectively.

## 2.2. SHG measurements

The most details of the SHG experiment have been described in the main text. A liquid-nitrogen cryostat (MicrostatHiRes, Oxford Instruments) was used for temperature-dependent measurements. Data collection at each temperature was performed after a stabilization of 5 minutes. The position of the sample is checked according to microscopy images during the measurements. Data analysis is performed using Igor Pro and Matlab software.

## 3. Results and discussions

### 3.1. Experiment setup and measurement protocols

We first briefly describe the design of our SHG microscope, as schematically shown in **Fig. 1a**. The excitation source is a Ti:sapphire mode-locking femtosecond laser (MaiTai SP, Spectra-Physics) generating a linearly polarized laser beam at 800 nm with a bandwidth of 30-60 nm. The polarization direction of the $1\omega$ light can be automatically rotated by a motorized half-wave plate before entering a pair of galvanometer mirrors. The latter mirrors couple with a $4f$ lens system and a 50× objective lens to scan the focused laser beam on the objective focal plane along both $x$ and $y$ axes. The objective focal plane can also be scanned along the $Z$ axis using a motorized stage. A normal angle of incidence is approximated only with significant deviations for large scanned sizes (*e.g.*, ~2° for 100 μm), which can be corrected for high-veracity data analysis. The generated $2\omega$ signal light is collected in the backscattering geometry and separated by a short-pass dichroic mirror. A Glan-Taylor prism polarizer, also automated by a motorized rotation stage, is used as an analyzer to allow transmission of SHG signals with certain polarization directions, which finally are photon-counted using a photomultiplier tube detector. Compared with those used in previous studies [24,36], our SHG microscope mainly differs in the adoption of a laser-scanning mode instead of a sample-scanning mode. The laser-scanning mode adds complexities to the optics setup, but brings in more flexibilities for sample environment apparatuses (such as multi-axial stages, cryostats and furnaces) thus remarkably facilitating spectral imaging measurements. The position repeatability errors of the galvanometer mirrors are verified to be below 100 nm and the long-time drifts of the microscope can be maintained below ~1 μm (see **Fig. S1**). These performance factors ensure that each pixel of an as-measured image sequence correspond to the same probed region within accuracies close to the optical diffraction limit.

For convenience, the samples are typically placed under the SHG microscope such that their crystallographic axes ($x$, $y$, $z$) coincide with the laboratory coordinate ($X$, $Y$, $Z$); the KNN crystals studied here are oriented along the pseudocubic <001>$_C$ axes. The setting of the half-wave plate angle $\alpha = 0°$ and analyzer angle $\varphi = 0°$ corresponds to the polarization directions of both 1$\omega$ and 2$\omega$ lights being parallel to the $X$ axis. A half-wave plate-analyzer angle map scan method is introduced to fully examine the possible SHG tensor components associated with different domains within the crystals; that is, the $\alpha$ and $\varphi$ angles are both adjusted stepwise from 0° to 360° to measure a complete map of SHG signals with differently combined polarization states. As shown in **Fig. 1b**, the obtained map clearly shows a two-fold symmetry with respect to both $\alpha$ and $\varphi$ angles. The major peaks marked in $\bigcirc$ are close to $\alpha = 90°$ (270°) and $\varphi = 90°$ (270°), which are found to align with the [±1 0 ±1] polarization vector (***P***) direction of the orthorhombic phase. The small deviations of ~10° in the peak positions are due to a miscut of the sample crystal and have been corrected in the following analysis. Meanwhile, minor peaks (marked in +) are observed near $\alpha = 0°$ (180°) and $\varphi = 90°$ (270°), indicating possible existence of a second domain variant which may belong to the same or a different phase.

One-dimensional (1D) spectra can be extracted from the $\alpha$-$\varphi$ angle scan map for standard SHG polarimetry analysis. **Fig. 1c** presents the polarization dependent SHG spectra for the analyzer angle $\alpha = 0°$ and 90°. These spectra have been fitted with different possible point groups of KNN ferroelectric phases (see **Fig. S2**), among which the orthorhombic *mm2* point group yields the best fit. On the other hand, the $\alpha$-$\varphi$ coupled scan spectrum (i.e., the diagonals in **Fig. 1b**) directly provide information about the in-plane anisotropy of the sample. As shown in **Fig. 1d**, the result can be well fitted by a model of two coexisting orthorhombic phase domain variants with the **P** directions of [±1 0 ±1]$_C$ (denoted as *O*1-domain) and [0 ±1 ±1]$_C$ (*O*2-domain), and based on their relative intensities, the fraction of the *O*1-domain is calculated to be ~73% within the probe sample volume.

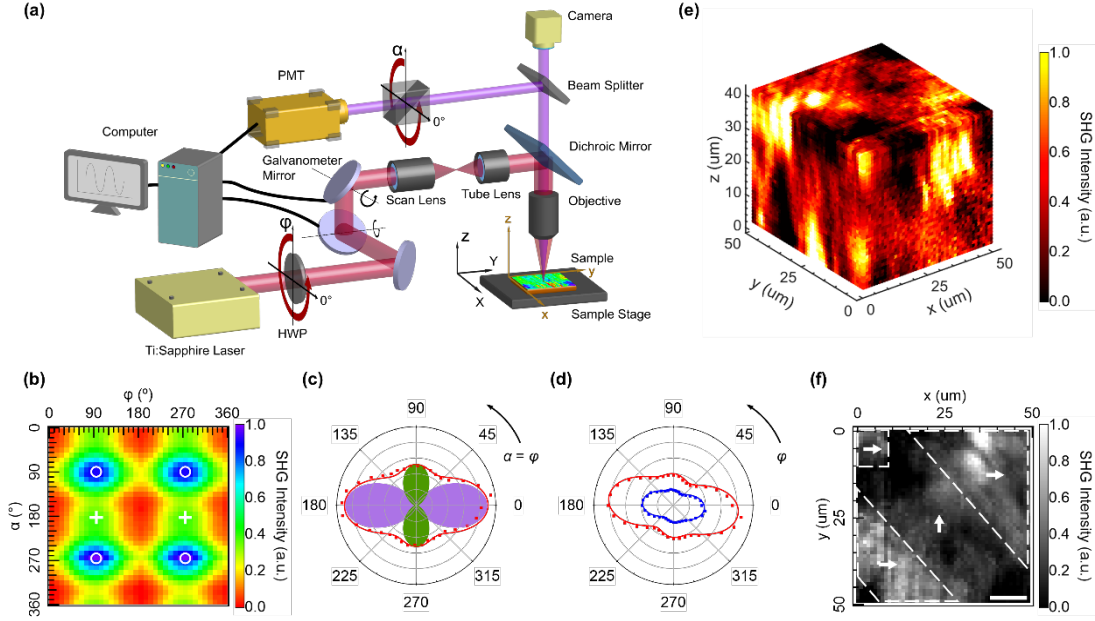

**Fig. 1.** The SHG microscopy setup and typical measurement results of KNN single crystals. **(a)** Schematic for the home-designed SHG microscope (not to scale). The visible-light illumination part is omitted for clarity. The $1\omega$ fundamental light and $2\omega$ SHG signal are colored in red and purple, respectively. (b) A complete SHG signal map for different measuring polarization states. The major peaks are marked in $\bigcirc$ and minor peaks marked in +. (c) The polarization dependent SHG spectra at $\alpha = 0°$ (red dots), $90°$ (blue dots) and corresponding fitting results (lines). (d) The $\alpha$-$\varphi$ angle coupled scan spectrum and corresponding fitting results. The two-fold profiles colored in purple and green refer to the contribution of the $O1$-domain and $O2$-domain, respectively. (e) Reconstructed 3D spatial distribution of SHG intensity and (f) the topmost image. Possible domain orientations of different regions are marked in arrows.

    **Fig. 1e** illustrates a typical 3D spatial distribution of SHG signals of the crystals, reconstructed from a set of focus-depth resolved 2D SHG images with the topmost image shown in **Fig. 1f**. Here, both $\alpha$ and $\varphi$ angles are fixed at 0°, for which conditions the measured intensity mainly originates from the $O1$-domain (see **Fig. 1d**). The observed intensity contrast thus reflects the orthorhombic phase domain configurations which, according to the crystallographic rules, features domain bands with domain wall boundaries running along the $[110]_C$ and $[101]_C$ directions. The domain contrast is not as homogeneous as expected for a simple two-level, $O1/O2$-domain configuration, suggesting a more complex and disordered domain microstructure. Furthermore, note that the observation of strong SHG signals across a large depth range is remarkable per se. This implies the coherent length of SHG within the KNN crystals are larger than the laser focus profile along the Z axis (~2 μm), which is not due to the intrinsic phase matching condition (*i.e.*, the difference in the refractive indices at 800 nm and 400 nm is rather small) but points to an incoherent domain structure of the

crystal (similar to the principle of quasi-phase matching). **Fig. S3** provides measurement results of a commercial high-quality LiNbO$_3$ single crystal for comparison.

### 3.2. SHG Spectral Imaging and Physical Model Fitting

To comprehensively reveal the heterogeneous domain configurations of the KNN crystals, we introduce a SHG spectral imaging method based on sequential imaging under various $\alpha$ and $\varphi$ angles. As illustrated in **Fig. 2a**, the SHG image stack consists of 36 images measured by varying $\alpha$ and $\varphi$ angles ($\alpha = \varphi$) within 0°-360° in step of 10°; each layer of the dataset can be individually analyzed as a traditional SHG image and each pixel of the dataset can be regarded as an anisotropy measurement spectrum. In-situ SHG spectral imaging measurements have been performed for the KNN single crystals over a temperature range of 80-480 K, and three representative datasets at 100 K, 280 K and 500 K are presented here with detailed analysis.

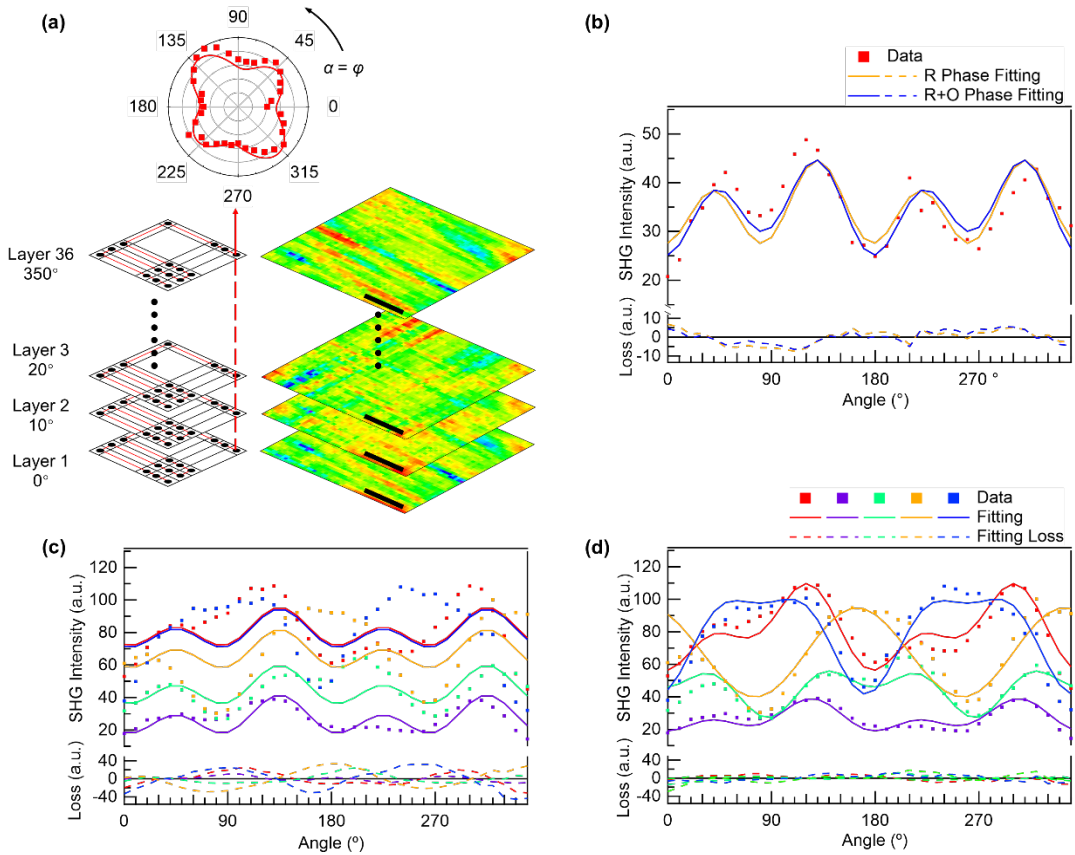

**Fig. 2.** The SHG spectral imaging method and physical model fitting. (a) Illustration for the measurement procedure showing several images and a compiled single-pixel spectrum. (b) Fitting results for a single-pixel spectrum of the 100 K data. A comparison between single *R*-phase and mixed *R/O*-phase models is made. (c) Global fitting result with the single *R*-phase model. A lack of variability of the trial function profile results in a poor fitting for some measured pixels. (d) Global fitting result with the mixed *R/O*-phase model showing good convergence.

The SHG spectral imaging data is first analyzed using a physical model fitting approach. The $R$-to-$O$ and $O$-to-$T$ phase transitions of KNN are believed to occur around 120 K and 460 K. To account for potential diffused transition behaviors, both the $R$ and $T$ phases are considered for the 100 K data while the $O$ and $T$ phases are considered for the 280 K and 500 K data. In general, the measured SHG signals of a mixed-phase sample can be fitted to a function F as follow:

$$F(\boldsymbol{d}, C, f_k, \theta) = \left( \sum_k \sum_{i,j} f_k d_{ij}^k E_i E_j \right)^2 \tag{2}$$

where $f_k$ denotes the fraction of a certain phase, $\boldsymbol{d}$ is the SHG tensor with coefficients $d_{ij}$ represented in Voigt notations, C is the instrument background and $\theta$ is the anisotropy measurement angle ($\theta = \alpha$ or $\varphi$). See **Table S1** for detailed function forms of the involved $R/O/T$- phases. Instead of fitting the spectrum of each individual pixel, a global fitting procedure has been performed. In the latter procedure, the total loss function is defined as the $L_2$ loss:

$$L(\boldsymbol{d}_0, \boldsymbol{d}_1, \dots, C) = \sum_{x,y} \sum_\theta |F(\boldsymbol{d}_0, \boldsymbol{d}_1, \dots, C, f, \theta) - Y(\theta)|^2 \tag{3}$$

where F is the trial fitting function described in **Eqn. 2** and Y is the measured SHG intensity. The individual losses are summed first over the anisotropy angle $\theta$ and then over all the pixels to give a global evaluation of the fitting result. Here, the SHG tensor $\boldsymbol{d}_i$ corresponding to the pixels and the background C are fixed, but the fractions of domain variants $f$ are allowed to vary in different regions of the sample.

The fitting results for a randomly selected pixel of the 100 K data are illustrated in **Fig. 2b**. The comparison between fixing $f_R = 1$ and not shows similar fitting goodness; that is, both single $R$-phase and mixed $R/O$-phase models can be compatible with the observed 4-fold symmetry pattern (note that a single $O$-phase model would produce 2-fold symmetry patterns similar to **Fig. 1d**). This is mainly caused by the many fitting parameters in the trial function leading to an overfitting of the single-pixel spectrum. By contrast, the global fitting procedure yields more well-defined results. As shown in **Fig. 2c and d**, the single $R$-phase model limits the variability of the trial function profile due to its single form of $\boldsymbol{d}$ tensor, leading to inadequate fitting goodness around 45° and 225°. Such limitation is lifted by the introduction of a mixed $O$-phase and good convergence is reached for all the pixels with a ~300 % lower total loss. Thus, the global fitting results clearly distinguish the existence of buried $O$-phase domains at this temperature.

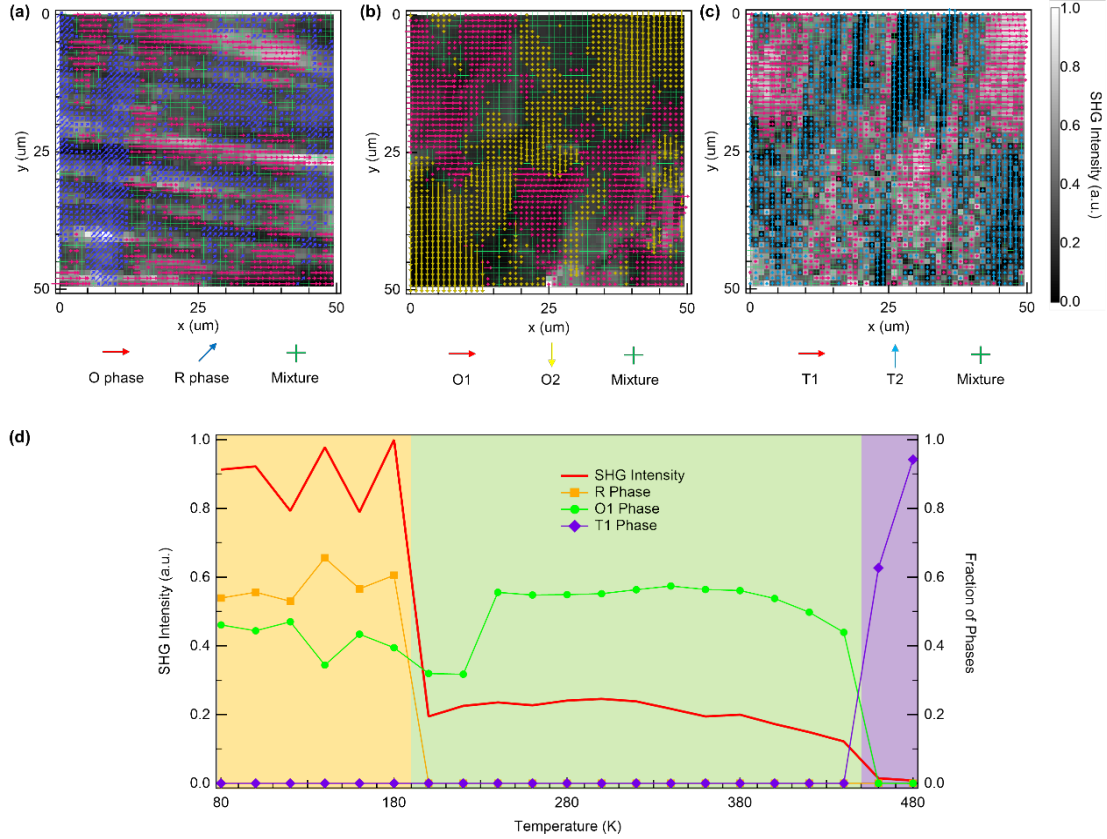

**Fig. 3.** Thermal evolution of domain configurations of KNN revealed by SHG spectral imaging. (a-c) Analyzed 2D maps of phase and domain variants for the same region measured at 100 K, 280 K and 500 K, respectively. The color and direction of arrows denote the major domain type for each pixel. The overlaid SHG images were measured at α = φ = 0°. (d) Temperature-dependent domain fractions and corresponding SHG intensity summer over the measured region.

As the global fitting results in phase/domain fraction values for each pixel, 2D maps can then be reconstructed for direct visualization of the domain configurations. These maps are presented in **Fig. 3a-c** overlaid with SHG intensity images measured at α = φ = 0°. Regions with a majority of *R*-phase and *O*-phase are marked with blue and red arrows, respectively, while regions with the two-phase fractions ~50% are marked with green cross. The 100 K map clearly shows coexisting stripe-like domains of the *O*-phase (***P*** along the $[\pm1\ 0\ \pm1]_C$) and *R*-phase (***P*** along the $[\pm1\ \pm1\ \pm1]_C$) patched with regions of the mixture. The latter regions may originate from an overlapping of the *O*- and *R*-phases within the probed single-pixel sample volume or equivalently, an intermediate crystal structure (likely the monoclinic phases) bearing both the *O*- and *R*-phase symmetry characters. The *O*-phase domain stripes exhibit high SHG intensities due to its large $d_{11}$ coefficient that dominates for the measuring polarization conditions. At 280 K, the crystals develop a somewhat uniform, single-phase domain structure consisting of the *O*1- and *O*2- domain variants (**Fig. 3b**). At 500 K, the SHG image again exhibits stripe patterns with a smaller average domain width (**Fig. 3c**). Two *T*-phase domain variants, *T*1- (***P*** along the [±1 0 0]$_C$) and *T*2- (***P*** along the [0 ±1 0]$_C$) domain, are identified and the *T*1-domain shows higher SHG intensities for the measuring polarization conditions. Notably, the domain stripe directions appear to be nearly normal to those displayed on the 100 K map. This may indicate a memory-like effect in the domain boundary reconstruction processes during the *R-O-T* phase transitions of KNN.

The continuous phase transition behavior of KNN over 80-480 K is then investigated using the above global fitting method. **Fig. 3d** presents the temperature-dependent SHG intensity and phase/domain fractions, both summed over the whole imaged region**.** The SHG intensity profile shows two abrupt drops around 200 K and 460 K, which apparently correspond to the *R-O* and *O-T* average structure transitions. The evolution of the phase/domain fractions overall agrees with the trend in the SHG intensity, and the occurrences of the *R*-phase and *T*-phase are observed only below 200 K and above 460 K, respectively. The most striking observation is the persistence of the *O*-phase down to the lowest measured temperature; over the imaged region, the *O*-phase fraction reaches nearly 50% below 200 K. It is well-known that the polymorphs of KNN have close free energies and their interconversions can be induced by applied electric field and mechanical stress. The stabilization mechanisms for the *O*-phase in the studied crystals can be due to residual stress arising from the crystal growth and/or sample preparation processes. A more intrinsic scenario can also be possible that the complex domain microstructure in KNN leads to strong influence on the phase stabilities in itself during the thermal evolution [37].

## 3.3. Unsupervised Matrix Decomposition Analysis

While the physical model fitting is a powerful method to extract detailed information from SHG measurements, a priori knowledge about the crystal phase symmetries of tested materials is required in order to apply correct forms of fitting function, and the fitting process itself needs close supervision. In this regard, unsupervised learning approaches can be advantageous to quickly, on-the-fly, decode the spectral imaging data, resulting in complementary and potentially more faithful interpretations of the obtained data. The efficacy of the latter approaches has been demonstrated for the analysis of a variety of microscopic and spectroscopic measurement results [38-40].

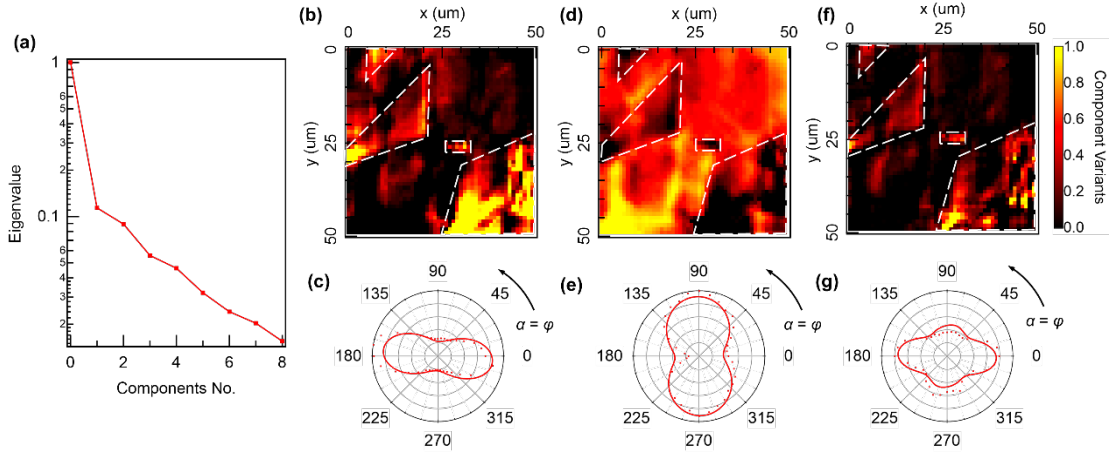

**Fig. 4.** Matrix decomposition analysis of SHG spectral imaging data. (**a**) Scree plot of variance ratio as a function of components number in the principle component analysis. (b,d,**f**) Component maps extracted from the original spectral imaging data by the three-component Bayes linear unmixing analysis and (c,e,g) corresponding SHG anisotropy spectra.

Here, we present the 280 K mapping data as an example, which consists of *O*-phase domains as discussed above. Principal component analysis (PCA) is first applied to determine the number components with statistically significant information. Here, the number of pixels of an SHG image is defined as the variable of feature space, while the length of each spectrum is defined as the number of samples, resulting in a $10000 \times 36$ data matrix. **Fig. 4a** presents the obtained scree plot of variance ratio, showing a sharp decrease from the first component to the second followed by another decrease from the second to the third. The components from the fourth on show a marginal decreasing trend with small variances. Therefore, the number of significant components is set to be three for the dataset. The complete PCA results are provided in **Fig. S4**. The PCA eigenvectors show rather complicated, unphysical spectral patterns which refrain us from further interpretation.

Instead, Bayes linear unmixing (BLU) is found to produce much more interpretable results, as presented in **Fig. 4b-g**. The obtained SHG anisotropy spectra from the three-component (the choice of the number is informed by the PCA analysis) BLU show well-defined patterns in **Fig. 4c,e,g** that can be associated with the *O*1-domain (***P*** along the $[\pm 1\ 0\ \pm 1]_C$), *O*2-domain (***P*** along the $[0\ \pm 1\ \pm 1]_C$) and their mixture (or likely a lower-symmetry monoclinic phase), respectively. Correspondingly, the component maps depict their spatial distributions as well as fraction ratios over the imaged region. The consistency between these maps and the model fitting results (**Fig. 3b**) can be confirmed by visual inspection. Interestingly, the maps for the *O*1-domain and the mixture region bear large resemblance (c.f. **Fig. 4b and f**); the spectral pattern of the latter is also dominated by the $[\pm 1\ 0\ \pm 1]_C$ polarization vector. This may suggest that the *O*1-domain is less stable than the *O*2-domain and prone to distortions or transformations in the studied KNN crystals. All told, these spectral imaging results reveal far more comprehensive information than SHG imaging or polarimetry alone.

## 4. Conclusions

In summary, we have expanded the SHG methodology by introducing the spectral imaging measurement and analysis method. Base on this method, we have found the co-existence of the *O*-phase and *R*-phase at low temperatures in KNN single crystals and continuously tracked the evolution of the constituent domain/phase fractions during heating. Quantitative delineation of the domain configurations is realized from combined physical model fitting and matrix decomposition analyses. These renewed insights of KNN may help devise ideas to enhance their ferroelectric and piezoelectric properties. With its efficacy proved through the model system, we recommend the spectral imaging to be a routine measurement for SHG studies of complex polar materials.

We further comment on several aspects for potential development. First, the spatial resolution of our current SHG microscope is at sub-micron levels due to the far-field diffraction limit. This can be improved by a factor of 1-2 using higher numerical-aperture optics, and near-field detection techniques, such as the tip-scattering method, can be incorporated to push into sub-100 nm levels. The resolution refinement can enhance delineation of complex structural distortions at domain and phase boundaries and may even allow effective probing of nanoscale polar textures. Second, more dimensions can be added into the measurement and analysis, such as focus depth and in-situ applied fields, to uncover correlated patterns or behaviors which otherwise are hidden in simplistic measurements. Finally, since SHG is a high-symmetry coupling effect, symmetry constraints can be introduced in the data analysis to boost efficiency and veracity, and more advanced data analytics, such as deep learning, should be explored.

## Conflict of Interest

The authors declare no conflict of interest.

## Acknowledgements

This work was supported by National Natural Science Foundation of China (NSFC) under Grants No. 52073155 and No. 52150092, and by the National Key Basic Research Program of China under

Grant No. 2020YFA0309100.

**Appendix A. Supplementary data**

Supplementary data to this article can be found online.